\documentclass[pra,aps,superscriptaddress,twocolumn]{revtex4}

\usepackage{epsfig}

\usepackage{bm}

\usepackage{color}

\newcommand{\petr}[1]{\textcolor{black}{#1}} 

\newcommand{\fr}[1]{\textcolor{black}{#1}} 
\newcommand{\jg}[1]{\textcolor{black}{#1}} 
\newcommand{\g}[1]{\textcolor{black}{#1}} 
\newcommand{\ex}[1]{\textcolor{black}{#1}} 

\newcommand{\Tr}{\mathrm{Tr}}

\begin{document}

\title{Multiple-copy distillation and purification of phase diffused squeezed states}

\author{Petr Marek}

\affiliation{School of Mathematics and Physics, The Queen's University,
  Belfast BT7 1NN, United Kingdom}

\author{Jarom\'{i}r Fiur\'{a}\v{s}ek}
\affiliation{Department of Optics, Palack\'y University, 17. listopadu 50, 77200 Olomouc,
Czech Republic}

\author{Boris Hage}
\affiliation{Max-Planck-Institut f\"ur Gravitationsphysik (Albert-Einstein-Institut)
and Leibniz Universit\"at Hannover, Callinstr. 38, 30167 Hannover, Germany}

\author{Alexander Franzen}
\affiliation{Max-Planck-Institut f\"ur Gravitationsphysik (Albert-Einstein-Institut)
and Leibniz Universit\"at Hannover, Callinstr. 38, 30167 Hannover, Germany}

\author{James DiGugliemo}
\affiliation{Max-Planck-Institut f\"ur Gravitationsphysik (Albert-Einstein-Institut)
and Leibniz Universit\"at Hannover, Callinstr. 38, 30167 Hannover, Germany}

\author{Roman Schnabel}
\affiliation{Max-Planck-Institut f\"ur Gravitationsphysik (Albert-Einstein-Institut)
and Leibniz Universit\"at Hannover, Callinstr. 38, 30167 Hannover, Germany}

\begin{abstract}
We provide a detailed theoretical analysis of multiple copy
  purification and distillation protocols for phase diffused
squeezed states of light. 
The standard iterative distillation 
protocol is generalized to a collective purification of an arbitrary number of $N$ copies. 
We also derive a semi-analytical expression for the asymptotic limit 
of the iterative distillation and purification protocol and discuss its properties.
\end{abstract}

\maketitle

\section{Introduction}
Continuous variable (CV) quantum information processing \cite{ContVar}
\g{is complementary to discrete variable quantum information  based on qubits.
 It employs the encoding of information into field modes of light
\cite{light1,light2,light3} or into the collective spin state of many particles such 
as in a cloud of atoms \cite{atom1,atom2,atom3}. 
Encoding quantum information into continuous variables of optical modes
exhibits several advantages over approaches based on  single-photon qubits. 
For example, it allows for the deterministic realization 
of several important protocols, such as the generation
of entanglement \cite{ent1,ent2,ent3,ent4} and entanglement swapping \cite{swap}, quantum teleportation
\cite{light1,light2}, quantum cloning \cite{clon}, and quantum dense coding \cite{dense}, using only linear
optics, optical parametric oscillators as squeezed light sources and balanced homodyne detection. It is also
possible to store continuous variables of light in atomic memories, either by means of direct interaction
\cite{atom2} or by teleportation \cite{atom3}.}

In the description of CV systems, an important role is played by probability quasidistributions,
which represent each field mode by a real two-dimensional function \cite{quasidistributions}. When
this function is of Gaussian shape, the corresponding state is referred to as \emph{Gaussian}. Due to
the ease of both experimental and theoretical treatment of such states, they are, together with Gaussian
operations preserving the Gaussian nature of the quasidistributions, a significant part of CV
quantum information. 
\petr{However, attempts to estabilish entanglement between two distant parties, which is one of the fundamental tasks quantum information is facing, 
are hampered by decoherence and thus, purification and distillation of entanglement are needed to remove the added noise and restore, at least approximately, the initial entangled state \cite{distill1,distill2}. Indeed, the entanglement distillation and purification techniques allow to extract from many copies of weakly entangled mixed state a single copy of highly entangled almost pure state by means of local quantum operations on each party's side and classical communication between the two parties sharing the states. In general, distillation refers to process of increasing entanglement (or squeezing), while purification is associated with reduction of the state's  mixedness.
Within the framework of this paper both effects involve each other and the corresponding terms will be used interchangeably.
}
Unfortunately, purification and distillation of entanglement are impossible to accomplish for 
Gaussian states by using feasible Gaussian operations alone \cite{nopurif1,nopurif2,nopurif3}. Note, that a similar
no-go theorem also holds for single mode squeezed states \cite{nopurifsq}. Namely,
it is impossible to use passive Gaussian operations, balanced homodyne detection and
feedforward to transform an arbitrary number of Gaussian states
into a single state with
squeezing better than the initial one, where the squeezing is characterized by
the lowest eigenvalue of the covariance matrix. On the other hand, Gaussian operations are sufficient to improve purity of a squeezed state if loss of squeezing is accepted \cite{distill5}.  

To allow for  increasing  the entanglement of Gaussian states by local means,
one has to at one point step out of the Gaussian domain, as was \g{utilized}
in \cite{distill3,distill4,Eisert1}. The complete distillation protocol,
proposed by \cite{Eisert1,Eisert2,Eisert3}, consists of two steps: de-Gaussification by
a single photon subtraction \cite{distill4,photonsub1,photonsub2,photonsub3,photonsub4}
followed by Gaussification by means of interference on beamsplitters, measurement
and conditioning. However, if the initial states are already non-Gaussian, for example
due to the effect of non-Gaussian noise, it is possible to distill and purify them
by employing only linear optics, homodyne detection and post-selection
\cite{nongdist1,nongdist2,nongdist3,nongdist4}. Recently, we have proposed \cite{nongdist1}
and experimentally demonstrated \cite{nongdist2, nongdist4} distillation of squeezing
from states disturbed by phase diffusion noise. This noise, caused by random fluctuations
of optical phase, commonly occurs in optical communication links. The distillation protocol
is based on the interference of two copies of the state at a balanced beamsplitter,
followed by homodyne detection of one mode triggering the acceptance or rejection of the
other mode that contains the purified state \cite{Eisert1,Eisert2,nongdist1}.
The procedure can be straightforwardly extended to accommodate two-mode squeezed states
and can also serve as  the Gaussification part of the universal CV distillation protocol based
on de-Gaussification and subsequent Gaussification.

In this paper we provide detailed theoretical numerical treatment of the distillation 
and purification of phase diffused single-mode squeezed states. 
We consider the standard iterative purification
 protocol \cite{nongdist1} as well as the simultaneous collective purification of an arbitrary number of copies
 of the decohered state.
We also discuss the possibility of conditioning on the measurement of an arbitrary quadrature operator, up to
abandoning the concept of definite quadrature completely and using phase-randomized homodyne detection.
Finally we derive a semi-analytical expression for the state obtained in the limit of
infinite number of iterations of the distillation procedure and discuss its
properties.

\begin{figure}
\centerline{\psfig{figure=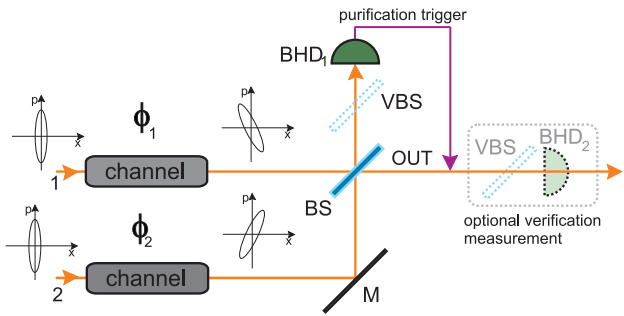,width=0.95\linewidth}}
\caption{\g{(Color online) Setup for purification of single-mode squeezed
  states. Two single-mode squeezed states are sent through independent
  dephasing channels which result in their decoherence.  They are then
  mixed on a balanced beamsplitter (BS) and homodyned in both
  output ports.  The virtual beamsplitters \petr{(VBS)} are meant to simulate
  inefficient homodyning.}} 
\label{setup1}
\end{figure}

\section{Iterative purification}
\label{iterativesection}
The single step of the purification protocol for single-mode squeezed states is depicted
in Fig.~\ref{setup1}. It consists of interference on a balanced \g{beamsplitter}
(BS) followed by
homodyne measurement of a single mode, detecting \fr{a} value of \fr{the} quadrature operator that was,
prior to effects of the noise, squeezed \cite{nongdist1,nongdist2}. If the absolute value of the measured quadrature $x$
is lower than a certain  pre-\fr{selected} threshold $X$ then the purification succeeded and otherwise
it failed. The purified state is present in the other output port of the \g{beamsplitter} and is available for
further applications or for \fr{the} next iteration of the purification procedure, whereby two purified
copies of the state are purified again.
The success of the procedure can be verified by suitable homodyne
measurements on the purified state, as schematically show\fr{n} in Fig. 1.
We will treat the purification in the Fock basis which
allows us to calculate the full density matrix of the
purified state after $k$ iterations of the purification protocol,
\begin{equation}
\rho^{(k)}= \sum_{m,n=0}^\infty \rho_{m,n}^{(k)}|m\rangle\langle n|,
\end{equation}
where $|n\rangle$ denotes the $n$-photon Fock state of the field mode.
The initial density matrix of a phase-diffused squeezed vacuum state can be expressed as
\begin{equation}
\rho^{(0)}=\int_\phi  U(\phi)\rho_G U^{\dagger}(\phi) \Phi(\phi) d \phi.
\label{rhozerodefinition}
\end{equation}
Here $\rho_{G}$ denotes density matrix of the initial
Gaussian state,  $\Phi(\phi)$  denotes the probability distribution of the random
phase shift $\phi$ and $U(\phi)$ is a unitary phase-shift operator with matrix elements 
$\langle m| U |n\rangle=e^{i n\phi}\delta_{mn}$, where the $\delta_{mn}$ stands for the Kronecker delta. We can rewrite Eq. (\ref{rhozerodefinition}) as follows,
\begin{equation}
\rho^{(0)}= \sum_{m,n=0}^\infty \rho_{G,m,n} f_{m-n} |m\rangle\langle n|,
\end{equation}
where $\rho_{G,m,n}=\langle m|\rho_G|n\rangle$ and
\begin{equation}
f_{n}= \int_{-\infty}^{\infty} \Phi(\phi) e^{in\phi} d \phi
\label{fk}
\end{equation}
is a Fourier transformation of $\Phi(\phi)$. Throughout this paper we assume that the phase
fluctuations have Gaussian distribution,
 \begin{equation}
 \Phi(\phi)= \frac{1}{\sqrt{2\pi \sigma^2}} e^{-\frac{\phi^2}{2\sigma^2}}.
 \end{equation}
On inserting this probability distribution into Eq. (\ref{fk}) we obtain 
$f_{n}=e^{-n^2\sigma^2/2}$. 

The Wigner function corresponding to the squeezed vacuum state $\rho_G$ reads
\begin{equation}
W(x,p)= \frac{1}{2\pi\sqrt{V_x V_p}} e^{-\frac{x^2}{2V_x}-\frac{p^2}{2V_p}},
\label{Wigner}
\end{equation}
where $V_x$ and $V_p$ denote the variances of \g{the} $x$ and $p$ quadratures, respectively. 
For a vacuum state one has $V_x=V_p=\frac{1}{2}$ and the state 
is squeezed when $V_x<\frac{1}{2}$ or $V_p < \frac{1}{2}$. 
\petr{Note that due to the proper normalisation, all variables used throughout the paper are dimensionless. 
Presently, these values could go as far as $V_x \approx 0.06$ - $0.05$ ($-9 dB$ to $-10 dB$), $V_p \approx 16$ - $25$ ($15 dB$ - $17 dB$) \cite{Furusawa, Schnabel}, but values around $V_x \approx 0.16$ - $0.22$ ($-5 dB$ to $-3 dB$) are common in contemporary experiments \cite{nongdist2,nongdist3,nongdist4}. 
}
The density matrix elements $\rho_{G,m,n}$  can be evaluated
 by noting that the Husimi Q-function defined as $Q(\alpha)= \frac{1}{\pi}\langle
\alpha|\rho|\alpha\rangle$, where $|\alpha\rangle$ denotes a coherent state with amplitude
$\alpha$, is a generating
function of the density matrix elements in \jg{the} Fock state basis,
\begin{equation}\label{densitym}
\rho_{m,n}=\frac{\pi}{\sqrt{m! \, n!}} \,\frac{\partial^{m+n}}{\partial \alpha^n
\partial \alpha^{\ast m}} \left.\left[e^{|\alpha|^2} Q(\alpha,\alpha^\ast)\right]
\right|_{\alpha=\alpha^\ast=0}.
\label{matrixelements}
\end{equation}
The Q-function corresponding to the Wigner function (\ref{Wigner}) is also Gaussian,
\begin{equation}
Q(\alpha,\alpha^\ast)= \frac{1}{\pi\sqrt{\tilde{V}_{x}\tilde{V}_p}}
\exp[(U-1)\alpha\alpha^\ast -T (\alpha^2+\alpha^{\ast 2})],
\label{Qfunction}
\end{equation}
where $\tilde{V}_x=V_x+1/2$, $\tilde{V}_{p}=V_p+1/2$ and
\begin{equation}
U=1-\frac{1}{2\tilde{V}_x}-\frac{1}{2\tilde{V}_p}, \qquad
T=\frac{1}{4\tilde{V}_x}-\frac{1}{4\tilde{V}_p}.
\end{equation}
On inserting function (\ref{Qfunction}) into expression (\ref{matrixelements})
we obtain after some algebra
\begin{equation}
\rho_{G,m,n}=\sqrt{\frac{m! \, n!}{\tilde{V}_x \tilde{V}_p}}
\sum_{a}\frac{(-T)^{\frac{m-n}{2}+2a} \, U^{n-2a}}{a! \,
(n-2a)!\, (a+\frac{m-n}{2})!}
\end{equation}
if $m+n$ is even and $\rho_{G,m,n}=0$ if $m+n$ is odd.

\begin{figure}[!t!]
\centerline{\psfig{figure=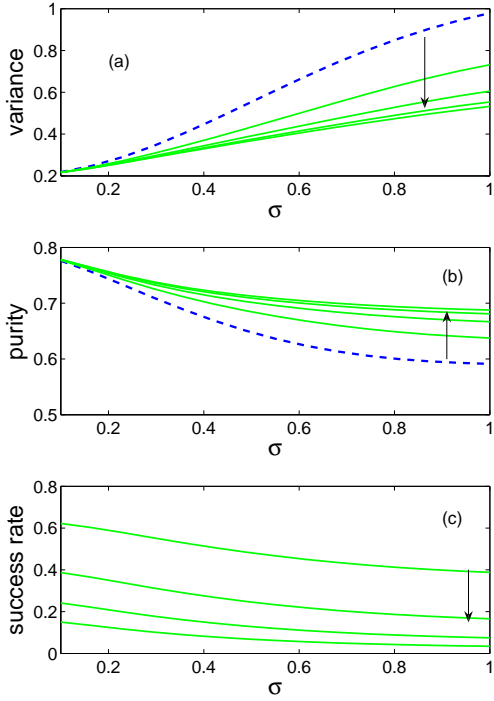,width=0.85\linewidth}}
\caption{\g{(Color online)} Performance of the four-\g{step} iterative purification protocol. The variance (a) and
 purity (b) of the purified state as well as the total success rate of the
 purification (c) are plotted as functions of the phase noise $\sigma$.
 The curve for the initial phase diffused state (blue dashed line) is plotted along curves
  of several iterations (green solid lines), where the arrows
 indicate the direction of increasing number of steps. The parameters were
  $\eta=0.85$, $V_x=0.2$, $V_p=2$, and $X=0.45$.}
\label{iterationsfig}
\end{figure}

The distillation and purification scheme shown in Fig.~1 produces with certain probability 
from two copies of the state $\rho^{(k-1)}$ a single copy of a purified state $\rho^{(k)}$
and we can formally write the purification map as
\begin{equation}
\rho^{(k)}= \mathcal{E}\left(\rho^{(k-1)}\otimes \rho^{(k-1)}\right).
\label{purificationmap}
\end{equation}
As the formula suggests, the protocol can be iterated and the outputs of each step 
can be used as inputs for the next iteration. As we shall show, each iteration increases the
squeezing and purity of the state and Gaussifies it. 

The interference of  two copies of the state $\rho^{(k-1)}$ on a balanced \g{beamsplitter} is
governed by a unitary transformation. Since this transformation preserves the
total photon number, we can write it in the Fock basis as follows,
\begin{equation}
|m_1,m_2\rangle \rightarrow \sum_{a} A_{m_1,m_2}^{a}|m_1+a,m_2-a\rangle,
\end{equation}
where
\ex{\begin{eqnarray}
A_{m_1,m_2}^a &=&\frac{\sqrt{m_1! \, m_2!}}{2^{(m_1+m_2)/2}} \nonumber \\
& & \times \sum_{d}\frac{(-1)^{d+a}\sqrt{(m_1+a)!\, (m_2-a)!}}{d!
(d+a)!(m_1-d)! (m_2-d-a)!}.
\nonumber \\
\end{eqnarray}}
After mixing on a BS the first output mode is measured in a balanced homodyne
detector where it is projected on the eigenstate of the $x$ quadrature $|x\rangle$.
The positive operator valued measure (POVM) element corresponding to conditioning 
on $|x|\leq X$ reads $C=\int_{-X}^X |x\rangle \langle x|$ and its matrix elements 
in \jg{the} Fock basis can be expressed as
\begin{equation}\label{postsel}
C_{m,n}= \int_{-X}^X  \langle m|x \rangle \langle x|n\rangle d x.
\end{equation}
This can be easily evaluated numerically by recalling that
\begin{equation}
\langle x|n\rangle = \frac{1}{\sqrt{\pi^{\frac{1}{2}} 2^n n!}} H_n(x) \, e^{-x^2/2},
\end{equation}
where $H_n(x)$ denotes the Hermite polynomial. For \fr{any} realistic detector with
efficiency $\eta <1$ we can use the model where \fr{the} inefficient detector is replaced
by a beamsplitter with transmittance $\eta$ whose auxiliary input
port is in the vacuum state and which is followed by an ideal perfect detector \petr{ \cite{quasidistributions}}. This yields
\begin{equation}
C_{m,n}(\eta)= \sum_{a} B_{m,a}(\eta)B_{n,a}(\eta)C_{m-a,n-a                                                                                },
\end{equation}
where
\begin{equation}
B_{m,a}=\sqrt{{m \choose a}} \eta^{(m-a)/2} (1-\eta)^{a/2}.
\end{equation}
We are now in a position to
combine all the above expressions and write down the purification map (\ref{purificationmap}) in the Fock
state basis,
\ex{\begin{eqnarray}
\rho^{(k)}& =&\sum_{m_1,n_1} \sum_{m_2,n_2} \sum_{a,b}
\rho_{m_1,n_1}^{(k-1)}\rho_{m_2,n_2}^{(k-1)} A_{m_1,m_2}^a A_{n_1,n_2}^b \nonumber \\
& & \times C_{m_1+a,n_1+b}(\eta) \, |m_2-a\rangle \langle n_2-b|. \nonumber \\
\label{rhoNplus1}
\end{eqnarray}}
The density matrix (\ref{rhoNplus1}) is not normalized and its trace is equal to the
probability of success \petr{(success rate)} of $k$ iterations of the purification protocol,
\begin{equation}
{P}(k)=\mathrm{Tr}[\rho^{(k)}].
\end{equation}

\begin{figure}[!t!]
\centerline{\psfig{figure=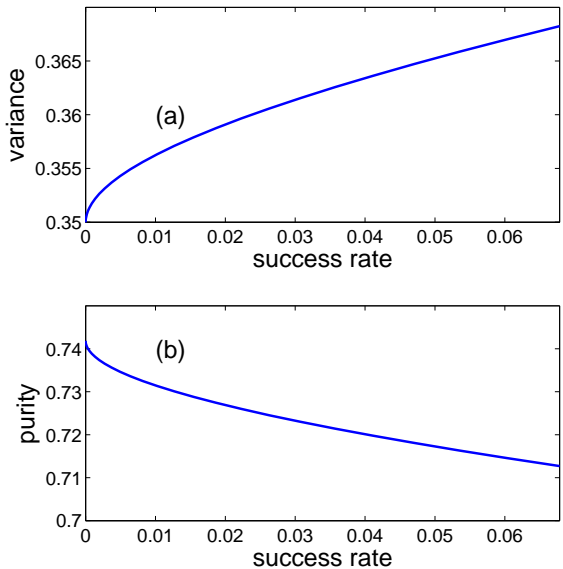,width=0.95\linewidth}}
\caption{\g{(Color online)} Trade-off between the success rate of the four-\g{step} purification protocol and
the resulting squeezing (a) and purity (b) of the state. \g{The
  parameters were  $\eta=0.85$, $V_x=0.2$,$V_p=2$, and $\sigma=0.5$.}}
\label{tradeofffig}
\end{figure}
\begin{figure}[!t!]
\centerline{\psfig{figure=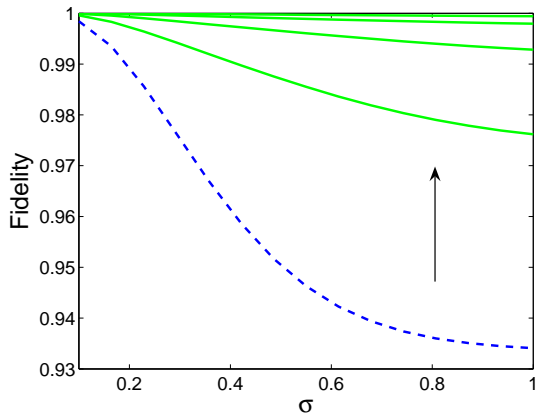,width=0.9\linewidth}}
\caption{\g{(Color online)} Fidelity of the purified state
with respect to a Gaussian state with the same covariance matrix
(green solid lines), the arrow indicates increasing number of
iteration steps. The lowermost curve (blue dashed line) corresponds to
the initial de-phased state. \g{The parameters were $\eta=0.85$, $V_x=0.2$, $V_p=2$, and $X=0.45$.}}
\label{fidelityfig}
\end{figure}

Typical numerical results are shown in Fig. \ref{iterationsfig}. We can see that each step
of the iterative purification decreases \fr{the} variance of $x$, i.e. increases the
squeezing. Also, the purity of the state
$\mathcal{P}=\mathrm{Tr}[\rho^2]/(\mathrm{Tr}[\rho])^2$ increases after each iteration of the protocol and
approaches some asymptotic value, which is generally less than unity, c.f. Fig
\ref{iterationsfig}(b).  So our protocol \jg{purifies the state but it does not generally
distill \g{perfectly} pure state\fr{s} from the initially mixed states}. \fr{One i}mportant parameter
determining the practical feasibility and usefulness of the scheme is the
probability of success of the protocol, which is plotted in Fig. \ref{iterationsfig}(c). We can
see that for the chosen acceptance window $X=0.45$ the success rate $P(N)$ is very high,
the probability of success of each iteration is about $50$\% and the total
probability of success for \fr{a} protocol involving four iterations is still
about $10$\%. The trade-off between the performance of the protocol and the
success probability is illustrated in Fig. \ref{tradeofffig} which shows the final variance and
purity of the state after four iterations as a function of the total success rate $P$. 
Higher purity and stronger squeezing are achievable at the expense of
reduced $P$. Our numerical calculations suggest that it is suitable to
choose the threshold $X \approx \sqrt{V_x}$, which \jg{achieves good
  purification and squeezing enhancement at a reasonably high probability of success.}

The purification procedure also Gaussifies the state. To quantify this, we
evaluate the fidelity of the state $\rho^{(k)}$ after $k$ iterations of the
protocol with the Gaussian state $\tilde{\rho}_G$ that has the same mean quadrature
values and covariance matrix as the state $\rho^{(k)}$. The fidelity of two
mixed states is defined as follows,
\begin{equation}
F= \left(\mathrm{Tr} \sqrt{\sqrt{\tilde{\rho}_G} \rho^{(k)} \sqrt{\tilde{\rho}_G}} \right)^2,
\end{equation}
and it holds that $F=1$ if and only if $\tilde{\rho}_G=\rho^{(k)}$. \fr{A v}alue $F<1$ is a clear signature of a
non-Gaussian character of the state $\rho^{(k)}$. The results are shown in Fig. \ref{fidelityfig} which
confirms that the present protocol indeed Gaussifies the state. After four iterations, we have $F>0.999$ so
the state is \g{almost} perfectly Gaussified.

\begin{figure}[!t!]
\centerline{\psfig{figure=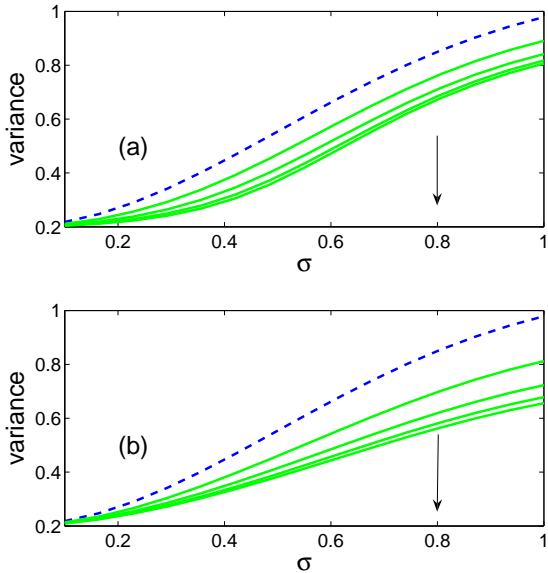,width=0.95\linewidth}}
\caption{\g{(Color online)} Iterative purification using conditioning on $|p|<Q$ (a) or
phase-randomized conditioning (b), \g{with the parameter settings} $\eta=0.85$, $V_x=0.2$, $V_p=2$, $Q=0.45$.
The arrows indicate increasing number of iterations (green solid lines).
The uppermost curve (blue dashed line) corresponds to the initial phase diffused state. }
\label{conjugateiterationsfig}
\end{figure}

So far we have considered only conditioning on \g{the} measurement of the initially squeezed
$x$ quadrature. However, the method works even for a measurement of \fr{an} arbitrary
quadrature $q(\theta) = x \cos\theta + p\sin\theta$ \cite{nongdist4}, where the
conditioning is again of the form $|q(\theta)| < Q$.  This can be straightforwardly
implemented into our Fock-state basis formalism by re-defining the matrix elements
of the POVM $C(\eta)$ as follows,
\begin{equation}
C_{m,n}(\eta,\theta)= C_{m,n}(\eta) e^{i(m-n)\theta}.
\label{Cmntheta}
\end{equation}
Note, however, that there are two prominent quadratures, $x = q(0)$ and $p = q(\pi/2)$.
Conditioning on one of these quadratures yields, for separate ranges of $\sigma$, 
the best improvement of the purified state. Furthermore, since improvement in the purified state can be seen
no matter which quadrature $q(\theta)$ is measured \cite{nongdist4}, it is possible
to consider also a phase-randomized homodyning. In this case, Eq. (\ref{Cmntheta})
has to be averaged over the random phase $\theta$ and only diagonal elements survive,
\begin{equation}
C_{m,n}^{\mathrm{rand}}(\eta)=C_{m,n}(\eta) \delta_{m,n}.
\end{equation}
The results of numerical simulations of the iterative purification are given in Fig. \ref{conjugateiterationsfig}. Note that the advantage of conditioning on $p$ grows with the number of iterations. For example, if $\sigma=0.5$ then after the first iteration the variance is more reduced by conditioning on $|x|<Q$ than on $|p|<Q$. However, for a four-step iterative procedure this is reversed and it is slightly better to condition on $|p|<Q$ rather than on $|x|<Q$. 
\petr{
The phase-randomised purification can be seen as a compromise between conditioning on $|x|<Q$ and $|p|<Q$ - for large (small) values of $\sigma$ it yields results better (worse) than conditioning on $p$ ($x$) but worse (better) than conditioning on $x$ ($p$). In fact, values obtained by phase-randomised purification are almost identical to those produced by random switching between $x$ and $p$ detections. 
}

\section{Collective purification}
\label{collectivesection}

\begin{figure}[!t!]
\centerline{\psfig{figure=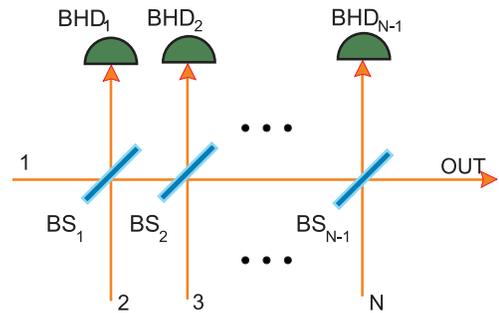,width=0.75\linewidth}}
\caption{\g{(Color online)} Collective purification of $N$ copies of the state. The beams are
combined on an array of $N-1$ \g{beamsplitters} BS$_{j}$ and all output modes except for
the last one are monitored with balanced homodyne detectors BHD$_{k}$.}
\label{collectiveschemefig}
\end{figure}

The iterative purification scheme analyzed in the previous section requires 
at least \ex{$2^k$} copies for \g{$k$} iterations of the protocol.
Here we analyze a more general collective purification scheme that works
for any number $N$ of available copies of the state. In particular, this protocol can be
employed when three copies of the state are available and it
\g{reduces} to the previous scheme when \ex{$N=2^k$}. The proposed setup is shown in Fig.
\ref{collectiveschemefig}. The $N$
copies of the phase-diffused squeezed state are combined on an array of $N-1$ \g{beamsplitters} with transmittances $t_j$ and reflectances $r_j$, $t_j^2+r_j^2=1$.
All output modes except for the last one are monitored by balanced homodyne
detectors which measure the $x$ quadrature of each mode. The purification is
successful if the absolute value of each measurement outcome is below the
corresponding threshold, $|x_{j,\mathrm{out}}| \leq X_j$. The choice of the thresholds
would depend on the transmittances of the beam splitters. Here we shall
consider ideal homodyne detectors with unit efficiency,
$\eta=1$, and conditioning on $|x_{j,\mathrm{out}}|=0$ for which the
protocol achieves the best performance. In practice, this would correspond to
choosing sufficiently narrow acceptance windows, i.e. very small thresholds $X_j$.

Let us for a moment fix the random phase shifts $\phi_j$ of each mode.
Then the variance of the quadrature $x_j$ of \fr{the} $j$-th mode reads
\begin{equation}
V_j=V_x \cos^2\phi_j+V_p \sin^2 \phi_j,
\end{equation}
and in what follows we shall not explicitly display the dependence of $V_j$ on $\phi_j$ for the
sake of brevity. The joint probability distribution of the $x$ quadratures of all
$N$ input modes \g{factors} into the product of marginal Gaussian probability
distributions,
\begin{equation}
P_{\mathrm{joint}}(\bm{x})= \prod_{j=1}^N \frac{1}{\sqrt{2\pi V_j}} e^{-\frac{x_j^2}{2
V_j}}.
\label{Pjointx}
\end{equation}
In the Heisenberg picture, the output $x$ quadratures after interference on the array of \g{beamsplitters} can be expressed as linear combinations of the input $x$ quadratures,
\begin{equation}
x_{j,\mathrm{out}}= \sum_{l=1}^N U_{j,l} x_l,
\label{inoutx}
\end{equation}
where $U_{j,l}$ denote elements of a real orthogonal matrix which describes the
action of the interferometer composed of the \g{beamsplitters}. For the scheme shown in
Fig. \ref{collectiveschemefig} it holds that
\begin{equation}
U_{N,l}=r_{l-1}\prod_{j=l}^{N-1} t_l.
\end{equation}
Here we formally define $r_0=1$. The input-output relation (\ref{inoutx}) 
can be inverted and we have
\begin{equation}
x_j=\sum_{l=1}^N (U^T)_{j,l}x_{l,\mathrm{out}}= \sum_{l=1}^N U_{l,j}x_{l,\mathrm{out}}.
\end{equation}
The joint distribution of the output quadratures can be obtained by inserting the
$x_j$ expressed in terms of $x_{l,\mathrm{out}}$ in Eq. (\ref{Pjointx}).
Taking into account that the Jacobian of the transformation (\ref{inoutx}) 
is $\det U=1$, we obtain
\begin{equation}
P_{\mathrm{joint}}(\bm{x}_{\mathrm{out}})= \prod_{j=1}^N \frac{1}{\sqrt{2\pi V_j}}
\exp\left[-\frac{1}{2 V_j} (\sum_{l=1}^N U_{l,j} x_{l,\mathrm{out}})^2\right].
\end{equation}
The un-normalized distribution of $x_{N,\mathrm{out}}$ conditional on
projections $x_{l,\mathrm{out}}=0$, $l=1,\ldots,N-1$ then reads
\begin{equation}
P_{\mathrm{cond}}(x_{N,\mathrm{out}}) \propto{\frac{ \sqrt{\tilde{V}}}{\prod_{j=1}^N  V_j^{1/2}}}
\frac{1}{\sqrt{2\pi  \tilde{V}}} e^{-\frac{x_{N,\mathrm{out}}^2}{2\tilde{V}}},
\end{equation}
where
\begin{equation}
\frac{1}{\tilde{V}}= \sum_{j=1}^N \frac{U_{N,j}^2}{V_j}.
\end{equation}
The variance of $x_{N,\mathrm{out}}$ can be calculated by averaging properly weighted
$\tilde{V}$ over the $N$ random phase shifts,
\begin{equation}
V_{\mathrm{out}}= \frac{1}{\mathcal{N}_N}
\int_{\bm{\phi}} \frac{\tilde{V}^{3/2}}{\prod_{l=1}^N V_l^{1/2}}
\prod_{j=1}^N \Phi(\phi_j) d \bm{\phi}.
\label{Vcollective}
\end{equation}
Here $d \bm{\phi}= d\phi_1 \cdots d \phi_N$ and $\mathcal{N}_N$ is a normalization constant,
\begin{equation}
\mathcal{N}_N =
\int_{\bm{\phi}} \frac{\tilde{V}^{1/2}}{\prod_{l=1}^N V_l^{1/2}}
\prod_{j=1}^N \Phi(\phi_j) d \bm{\phi}.
\label{Ncollective}
\end{equation}
Two choices of the \g{beamsplitting} ratios are of particular interest. Using
balanced \g{beamsplitters} is experimentally most straightforward, 
since these \g{beamsplitters} are most commonly used in the lab. In this case we have
$t_j=r_j=2^{-1/2}$ which implies  $U_{N,1}=2^{-(N-1)/2}$ and
$U_{N,l}=2^{-(N-l+1)/2}$, $l\geq 2$. Notice that the output quadrature $x_{N,\mathrm{out}}$
is an unbalanced combination of the input quadratures.
The second choice is to use unbalanced \g{beamsplitters} which yield a balanced
superposition of the input quadratures, $U_{N,l}=1/\sqrt{N}$.  This happens
if $t_j=\sqrt{j/(j+1)}$. This latter scheme becomes fully equivalent to the $k$-step
iterative purification procedure discussed in the previous section when \g{$N=2^k$}.
Note that in the iterative scheme we can also first let all the beams interfere on
the balanced  \g{beamsplitters}  and then measure on all output ports except one.
If we condition on $x_{l,\mathrm{out}}=0$ then the two schemes become immediately fully
equivalent. For realistic finite thresholds $X_j$ the schemes become equivalent if
conditioning on appropriate linear combinations of $x_{l,\mathrm{out}}$ is adopted.

\begin{figure}[!t!]
\centerline{\psfig{figure=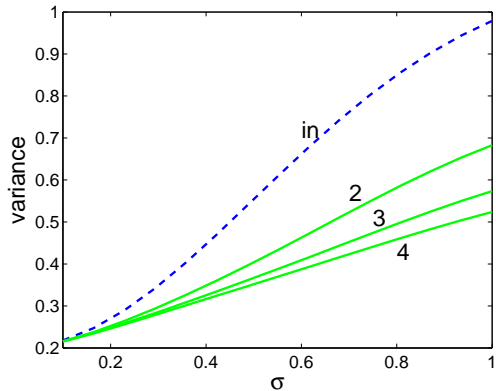,width=0.85\linewidth}}
\caption{\g{(Color online)} Variance of \g{the} $x$ quadrature of the state obtained by purification from $2$, $3$, and $4$
copies is plotted in dependence on $\sigma$. The curve labeled ``in'' shows
the variance for the initial state. The initial parameters were $V_x = 0.2$, $V_p = 2$, $\eta = 1$ and $X=0$.}
\label{collectivefig}
\end{figure}
\jg{The results} of numerical simulations of the collective purification procedure are
shown in Fig. \ref{collectivefig}. We can see that each additional copy helps to further suppress
the noise and reduce the quadrature fluctuations. The difference between the
performances of the two-copy and three-copy protocols is sufficiently large so that it
should be possible to observe it in present-day experiments. The numerics also
reveal that the difference between the schemes using balanced and unbalanced \g{beamsplitters} is very small and the scheme is quite robust and insensitive to changes
in reflectances or transmittances of the \g{beamsplitters} of the order of a few
percent, which is an important practical advantage.

\g{A conditional measurement} in the form of \g{a} detection of arbitrary quadrature $q(\theta)$
can be also considered for the collective purification.
To treat in a unified way conditioning on measurements of arbitrary quadratures of the
first $N-1$ output modes, we shall present a more detailed treatment of the scenario, based
on the formalism of covariance matrices (CM).
Let $\xi=(x_1,p_1,x_2,p_2,\ldots,x_N,p_N)$ denote the vector of quadrature
operators. The covariance matrix $\Sigma_{jk}=\frac{1}{2}\langle\{\Delta \xi_j,\Delta \xi_{k}\}\rangle$
comprises the second moments. The CM of the single-mode squeezed state is
diagonal, $\Sigma_{\mathrm{SMS}}=\mathrm{diag}(V_x,V_p)$. A random phase shift $\phi$
transforms \jg{the} CM to $\Sigma_{\mathrm{SMS}}(\phi)=R(\phi)\Sigma_{\mathrm{SMS}}R^T(\phi)$, where
\begin{equation}\label{rotace}
R(\phi)= \left(
\begin{array}{cc}
\cos\phi  & -\sin \phi \\
\sin \phi & \cos \phi
\end{array}
\right).
\end{equation}

\begin{figure}[!t!]
\centerline{\psfig{figure=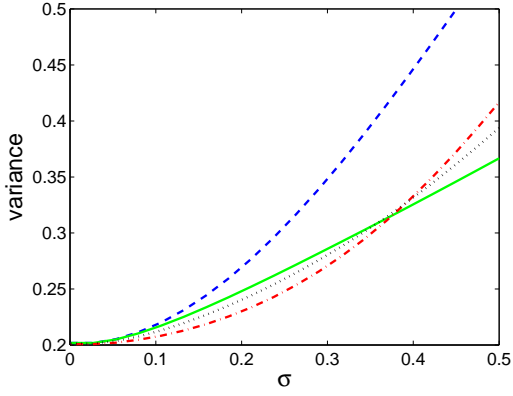,width=0.85\linewidth}}
\caption{\g{(Color online)} Collective purification from three copies of the state. The figure shows
the variance  of the $x$ quadrature before (blue dashed line) and after purification.
The quadratures measured by balanced homodyne detectors BHD$_1$ and BHD$_2$ were the following: $x_1$, $x_2$ (green solid
line), $p_1$, $p_2$ (red dot-dashed line), $x_1$, $p_2$ (black dotted line).}
\label{conjugatecollectivefig}
\end{figure}
For fixed  random phase shifts $\phi_j$, the covariance matrix of the N-mode
Gaussian state impinging on the array of $N-1$ \g{beamsplitters} is given by
\begin{equation}
\Sigma_{\mathrm{in}}= \bigoplus_{j=1}^N \Sigma_{\mathrm{SMS}}(\phi_j).
\end{equation}
The covariance matrix of the output modes can be obtained from $\Sigma_{\mathrm{in}}$ by
a symplectic transformation,
\begin{equation}
\Sigma_{\mathrm{out}}= S_{\mathrm{BS}}\Sigma_{\mathrm{in}}S_{\mathrm{BS}}^T,
\end{equation}
where the symplectic  matrix $S_{\mathrm{BS}}$ governs the linear mixing of
the quadrature components in the passive linear interferometer consisting of
the array of $N-1$ \g{beamsplitters}, see Fig. \ref{collectiveschemefig}. To make our
treatment simple, we shall again consider the limiting case of conditioning
on $q_{j}=0$, $j=1,\ldots,N-1$. The quadratures which are being observed are specified
by $N-1$ relative phases $\theta_j$ and we assume that all detectors have the same
efficiency $\eta$. We introduce a new covariance matrix
\begin{equation}
\Sigma_{\mathrm{aux}}= \eta S(\bm{\theta}) \Sigma_{\mathrm{out}} S^T(\bm{\theta})+\frac{(1-\eta)}{2}I,
\end{equation}
where $I$ stands for the identity matrix, $S(\bm{\theta})=\bigoplus_{j=1}^N
R(\theta_j)$ and we set $\theta_N=0$. The single-mode phase shifts in the above
formula map all measured quadratures onto the $x$ quadratures. We want to
calculate the variance of the quadrature $x_N$ conditional on all measurement
outcomes being equal to $0$. We first construct a submatrix of $\Sigma_{\mathrm{aux}}$ which
comprises variances  and covariances of the $x$ quadratures only,
\begin{equation}
\Sigma_{x}= S_x \Sigma_{\mathrm{aux}} S_x^T.
\end{equation}
Here $S_x$ is a $N \times 2N$ matrix defined as $S_{j,2j-1}=1$ and all other
elements are equal to zero. The joint probability distribution of the $N$
quadratures $\bm{x}=(x_1,x_2,\ldots,x_N)$ can be expressed as
\begin{equation}
P(\bm{x})= \frac{1}{(2\pi)^{N/2}\sqrt{|\Sigma_x|}} e^{-\frac{1}{2}\bm{x}^T \Sigma_{x}^{-1}\bm{x}},
\end{equation}
where the $|.|$ denotes determinant of the matrix.
The probability distribution of $x_N$ conditional on $x_1=x_2\ldots=x_{N-1}=0$ reads
\begin{equation}
P_{\mathrm{cond}}(x_N) \propto \sqrt{\frac{V_N}{|\Sigma_x|}} \frac{1}{\sqrt{2 \pi V_N}}
e^{-\frac{x_N^2}{2V_N}},
\end{equation}
where $V_N=1/[(\Sigma_{x}^{-1})_{NN}]$ is the conditional variance of $x_N$.
Since we assume that the mean values of all quadratures are initially zero and the
purification scheme preserves this property, \ex{the resulting variance of $x_N$
after purification can be evaluated as properly weighted average of $V_N$,}
\begin{equation}
V_{\mathrm{out}}= \frac{1}{\mathcal{N}_N}
\int_{\bm{\phi}}  V_N \sqrt{\frac{V_N}{|\Sigma_x|}}
\prod_{j=1}^N \Phi(\phi_j) d \bm{\phi}.
\label{Voutqcol}
\end{equation}
Here $d \bm{\phi}= d\phi_1 \cdots d \phi_N$ and $\mathcal{N}_N$ is a normalization constant,
\begin{equation}
\mathcal{N}_N =
\int_{\bm{\phi}} \sqrt{\frac{V_N}{|\Sigma_x|} }
\prod_{j=1}^N \Phi(\phi_j) d \bm{\phi}.
\label{Nqcol}
\end{equation}
Note that $V_{\mathrm{out}}$ determined in this way corresponds to the variance of the purified
beam that would be measured by balanced homodyne detector with efficiency $\eta$.
The expressions (\ref{Voutqcol}) and (\ref{Nqcol}) generalize the formulas (\ref{Vcollective}) and (\ref{Ncollective}).
\g{The} numerical results for three-copy purification are plotted
in Fig. \ref{conjugatecollectivefig}. This figure  confirms
the general trend that for weak phase fluctuations it is advantageous
to condition on measurements of $p$ while for strong noise it is better to
measure the quadrature $x$ \cite{nongdist4}. The intermediate strategy where the first balanced homodyne detector (BHD) measures
$x$ while the second BHD measures $p$ does not bring any advantage so we can
conclude that the optimum strategy that provides maximum reduction of quadrature
variance consists of measuring the same quadrature (either $x$ or $p$) by both
homodyne detectors.

\section{Asymptotic limit}
\label{sectionasymptotic} %
\begin{figure}[!t!]
\centerline{\psfig{figure=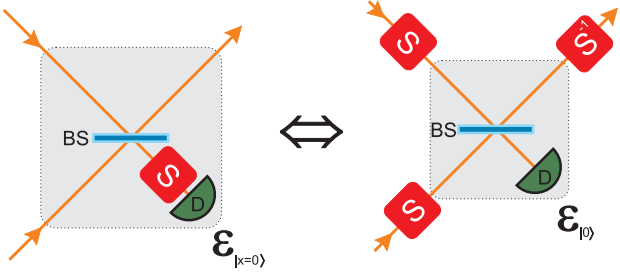,width=0.95\linewidth}}
\caption{\g{(Color online)} Equivalence of \g{the} single-step of the purification for homodyning (left)
and projection onto vacuum (right) as a conditioning measurement.}
\label{singlestepeq}
\end{figure}

\begin{figure}[!t!]
\centerline{\psfig{figure=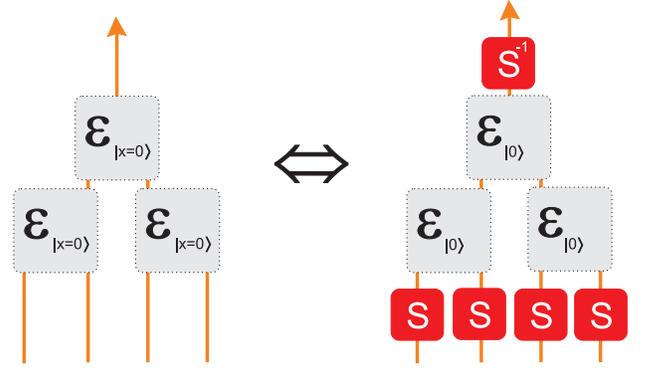,width=0.95\linewidth}}
\caption{\g{(Color online)} Equivalence of \g{the} iterative purification setup for homodyning (left)
and projection onto vacuum (right) as a conditioning measurements.}
\label{multistepeq}
\end{figure}

It is possible to find a state to which the purification procedure (\ref{purificationmap})
converges in the limit of infinite number of iterations, $k\rightarrow \infty$. This asymptotic state
can serve as a reference for the performance of the purification,
giving values that can never be surpassed by finite number of iterations.
Due to \jg{the} nature of the result we seek, we shall consider the post-selection
threshold $X$ in Eq. (\ref{postsel}) to be zero. That is, given
two copies of \fr{a} noisy state $\rho_{\mathrm{in},j}$, $j=1,2$,
the state after a single step of the purification is given by the map
\begin{eqnarray}\label{asympt1}
  \rho_{\mathrm{out}} &=& \mathcal{E}_{|x=0\rangle}(\rho_{\mathrm{in},1}\otimes\rho_{\mathrm{in},2}) \nonumber \\
    &=& \Tr_2[
    U_{12}\rho_{\mathrm{in},1}\otimes\rho_{\mathrm{in},2}U_{12}^{\dag}
    |x=0\rangle_2\langle x=0|],
\end{eqnarray}
where $U_{12}$ denotes \fr{a} unitary operator
describing balanced \g{beamsplitter} coupling modes $1$ and $2$ and $|x=0\rangle_2\langle
x=0|$ is \fr{a} projector on the eigenstate of the $x$ quadrature. In the
following we shall exploit the possibility of expressing the $x$
eigenstate as
\begin{equation}
|x=0\rangle = \lim_{r\rightarrow \infty} S_{r}
|0\rangle,
\end{equation}
where the operator $S_r = \exp[r(a^2-a^{\dag 2})/2]$ stands for
finite squeezing with parameter $r$ and $|0\rangle$ is the vacuum state \cite{Eisert3}.  
Since \g{the} operator $S_{r,1}\otimes S_{r,2}$ commutes with the \g{beamsplitter} transformation $U_{12}$,
we can rewrite the relation (\ref{asympt1}) as
\begin{eqnarray}\label{asymptotics2}
    \rho_{\mathrm{out}} &=& \lim_{r\rightarrow \infty}
    S_{r}\mathcal{E}_{|0\rangle}(\rho'_{\mathrm{in},1r}\otimes\rho'_{\mathrm{in},2r})S_{r}^{\dag}   \nonumber \\
    &=& \lim_{r\rightarrow \infty} S_{r} \langle 0|_2
    U_{12} \rho'_{\mathrm{in},1r}\otimes \rho'_{\mathrm{in},2r}
    U_{12}^{\dag}|0\rangle_2 S_{r}^{\dag}
\end{eqnarray}
with $\rho'_{\mathrm{in},jr} = S_{r}^{\dag}\rho_{\mathrm{in},j}S_{r}$ and where
the purification map $\mathcal{E}_{|0\rangle}$ is the same as in
Ref. \cite{Eisert2}. This is schematically demonstrated in
Fig.~\ref{singlestepeq}. Note
that the inverse squeezing operation performed on the output state
cancels out with the initial squeezing of the next purification
step. Thus, the iterative purification scheme, using projection
\g{onto the} quadrature eigenvectors in all steps, is equivalent to
iterative purification with projection \g{onto} vacuum states,
accompanied by squeezing of all initial states and inverse
squeezing of the purification outcome \cite{Eisert3}, see Fig.~\ref{multistepeq}. This allows us to find the
asymptotic limit, analogously as in Ref. \cite{Eisert2}. Due to this
similarity, we shall mention here only the most important parts of the
derivation and the reader can find more details
in Ref. \cite{Eisert2}.


The key realization in finding the limit is that stationary points
of the purification map
$\mathcal{E}_{|0\rangle}(\rho\otimes\rho)$, i.e. states that
remain unchanged by the map, are Gaussian states with zero mean
value. The initial non-Gaussian state will, over the course of
purification, converge to one of these states. As another
important observation, let us note that \g{the} coefficients
$\sigma_{11},\sigma_{20}$ and $\sigma_{02}$ defined as
$\sigma_{ij} = \rho_{ij}/\rho_{00}$ are also unchanged by the
purification map $\mathcal{E}_{|0\rangle}$ and there is a
one-to-one correspondence between these three coefficients and the set
of covariance matrices, which in turn completely describe any
single-mode Gaussian state with zero mean.

The procedure of finding the limit is the following: first, the
initial states are described by the Wigner function
\begin{equation}
W(x,p) = \int \Phi(\phi)
    \frac{1}{2\pi\sqrt{|\Sigma_{r,\phi}|}}\exp\left(-\frac{1}{2}\xi\Sigma_{r,\phi}^{-1}\xi^T\right) d\phi,
\end{equation}
where the vector $\xi = (x,p)$ comprises the two conjugate quadratures and
\begin{eqnarray}
    \Sigma_{r,\phi} &=& S_r R(\phi) \left(%
\begin{array}{cc}
  V_x & 0 \\
  0 & V_p \\
\end{array}%
\right) R^T(\phi) S_r,
    \nonumber \\
     S_r &=& \left(%
\begin{array}{cc}
  e^{r} & 0 \\
  0 & e^{-r} \\
\end{array}%
\right)
\end{eqnarray}
and $R(\phi)$ is defined  in Eq. (\ref{rotace}).
The Q-function corresponding to this Wigner function reads
\begin{equation}
    Q(x,p) =\int \Phi(\phi)
    \frac{\sqrt{|\Gamma_{r,\phi}|}}{2\pi}\exp\left(-\frac{1}{2}\xi\Gamma_{r,\phi}\xi^T\right) d\phi,
\end{equation}
with $\Gamma_{r,\phi} = (\Sigma_{r,\phi} + I/2)^{-1}$, where
$I$ stands for the identity matrix. Now, with the substitution $x =
(\alpha+\alpha^*)/\sqrt{2}$, $p = (\alpha-\alpha^*)/i\sqrt{2}$ we
can use relation (\ref{densitym}) and find the density matrix
elements
\begin{eqnarray}\label{sigma1}
\rho_{00} &=& \left\langle \sqrt{|\Gamma_{r,\phi}|} \right\rangle_{\phi} \nonumber \\
\rho_{11} &=& \left\langle \sqrt{|\Gamma_{r,\phi}|} (1 - \frac{\Gamma_{r,\phi;11}+\Gamma_{r,\phi;22}}{2})\right\rangle_{\phi}\nonumber \\
\rho_{02} &=& \left\langle \sqrt{|\Gamma_{r,\phi}|}\frac{\Gamma_{r,\phi;22}-\Gamma_{r,\phi;11}+2i\Gamma_{r,\phi,21}}{2\sqrt{2}}\right\rangle_{\phi}\nonumber \\
\rho_{20} &=& \left\langle \sqrt{|\Gamma_{r,\phi}|}\frac{\Gamma_{r,\phi;22}-\Gamma_{r,\phi;11}-2i\Gamma_{r,\phi,21}}{2\sqrt{2}}\right\rangle_{\phi},
\end{eqnarray}
where $\langle\rangle_\phi$ denotes averaging over the distribution $\Phi(\phi)$.
From the density matrix elements we can obtain the coefficients
$\sigma_{11},\sigma_{20}$ and $\sigma_{02}$, $\sigma_{ij} = \rho_{ij}/\rho_{00}$,
which are invariant over the course of the purification and, therefore, describe the
Gaussian limit \g{to which the state converges}. It is now a simple matter
of finding the inverse relations
\begin{eqnarray}\label{sigma2}
  \Gamma_{11} &=& 1-\sigma_{11}-\frac{\sigma_{02}+\sigma_{20}}{\sqrt{2}}, \nonumber \\
  \Gamma_{22} &=& 1-\sigma_{11}+\frac{\sigma_{02}+\sigma_{20}}{\sqrt{2}}, \nonumber \\
  \Gamma_{21} &=& \Gamma_{12} =
  \frac{\sigma_{02}-\sigma_{20}}{i\sqrt{2}}.
\end{eqnarray}
After some algebra we find that the formulas (\ref{sigma1}) and (\ref{sigma2}) 
can be simplified to
\begin{equation}
\Gamma = \frac{\langle \Gamma_{r,\phi}\sqrt{|\Gamma_{r,\phi}|}\rangle_{\phi}}{\langle
\sqrt{|\Gamma_{r,\phi}|}\rangle_{\phi}}.
\end{equation}
The final covariance matrix is given by $\Sigma =
S_r^{-1}(\Gamma^{-1} - I/2)S_r^{-1}$. Due to the symmetric nature of the phase noise,
the matrix $\Sigma$ is diagonal and fully specified by values of variances of the squeezed and
antisqueezed quadratures, $V_{x,\mathrm{lim}}$ and $V_{p,\mathrm{lim}}$.
We can directly find the output variances by taking the limit $r\rightarrow
\infty$, obtaining
\begin{eqnarray}
V_{x,\mathrm{lim}} &=& \frac{\langle A\rangle_{\phi}}{\langle A^3\rangle_{\phi}},\nonumber \\
V_{p,\mathrm{lim}} &=& V_x V_p \frac{\langle A^3\rangle_{\phi}}{\langle A\rangle_{\phi}},
\label{Vlim}
\end{eqnarray}
where
\begin{equation}
A = (V_x \cos^2\phi + V_p \sin^2 \phi)^{-1/2}.
\end{equation}
Note that $V_{x,\mathrm{lim}}V_{p,\mathrm{lim}}=V_xV_p$ so  the purification procedure
asymptotically suppresses all added noise and the asymptotic Gaussian state has the same purity
$\mathcal{P}=1/(2\sqrt{V_xV_p})$ as the initial state before de-phasing. This, however, holds only
for idealized conditioning on $x=0$. For finite acceptance window, $X>0$, the purity of the
asymptotic distilled state would generally be lower than the purity of the initial state.

\begin{figure}[!t!]
\centerline{\psfig{figure=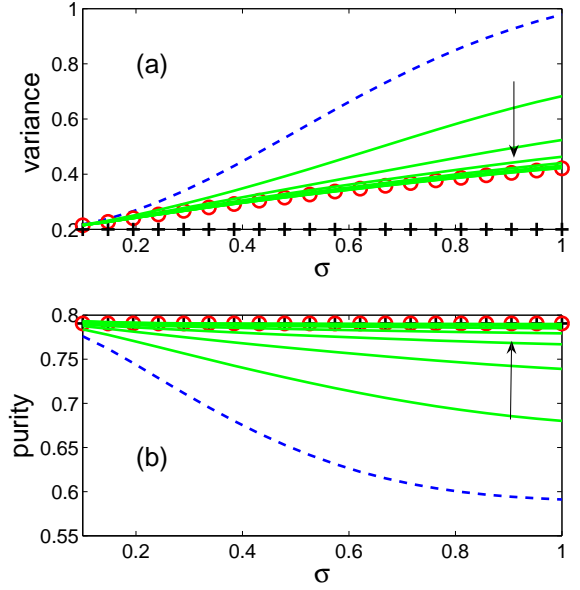,width=0.95\linewidth}}
\caption{\g{(Color online)} Performance of the iterative purification protocol (green line), compared to the initial phase diffused state (blue dashed line), asymptotic limit (red circles) and state without phase diffusion (black plus signs). The arrows indicate increasing number of iterations. The values shown are as if verified by an imperfect detector. The parameters were $V_x = 0.2$, $V_p = 2$, $\eta = 0.85$ and $X=0$.  }
\label{limit}
\end{figure}

It is possible to modify the method to incorporate other scenarios discussed
in previous sections. Finding the limit state for iterations incorporating conditioning
measurements along an arbitrary quadrature $q(\theta)$ is straightforward, requiring only
the use of \g{an} alternative squeezing operation described by $S_{r,\theta} = R(\theta)S_r R^T(\theta)$.
Adapting the calculations to incorporate for imperfect detectors is a bit more tricky.
Inefficient detection with efficiency $\eta$ can be simulated by a \g{beamsplitter}
with transmittance $\eta$ followed by a perfect detector. The attenuation can be represented
by a map,  $\mathcal{M}(\rho)$, acting on the density matrix and transforming the corresponding
covariance matrix of the Gaussian state as,
\begin{equation}
\Sigma \rightarrow \eta \Sigma   + \frac{1-\eta}{2} I.
\end{equation}
The inverse operation, $\mathcal{M}^{-1}(\rho)$ acting as
\begin{equation}\label{Ninv}
\Sigma \rightarrow \frac{1}{\eta}( \Sigma  - \frac{1-\eta}{2} I)
\end{equation}
is, of course, nonphysical. We can, however, formally 
use it and in analogy with Eq. (\ref{asymptotics2})
represent \fr{a} single step of \fr{the} purification procedure by \fr{the} relation
\begin{eqnarray}\label{noisystep}
\rho_{\mathrm{out}} &=& \mathcal{M}^{-1}\left(\lim_{r\rightarrow\infty} S_r \varepsilon S_r^{\dag}\right),
\end{eqnarray}
where the $\varepsilon$ is given by 
\begin{eqnarray}
\varepsilon &=&  \mathcal{E}_{|0\rangle}\left(S_{r,1}^{\dag}\mathcal{M}(\rho_{\mathrm{in},1})S_{r,1}\otimes
S_{r,2}^{\dag}\mathcal{M}(\rho_{\mathrm{in},2})S_{r,2}\right).
\end{eqnarray}
Now again, starting from \g{the} modified input states $S_r^{\dag}\mathcal{M}(\rho) S_r$
we can find the asymptotic limit for the map $\mathcal{E}_{|0\rangle}$ and with
the use of relation (\ref{noisystep}) \g{it can be transformed into
  the desired result}. However, if we are interested in result\fr{s} that can be observed by imperfect detection,
the final operation $\mathcal{M}^{-1}$ is not necessary.

\begin{figure}[!t!]
\centerline{\psfig{figure=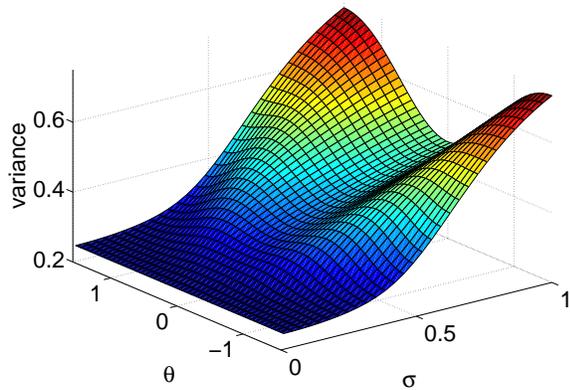,width=0.95\linewidth}}
\caption{\g{(Color online)} Variance of the asymptotic limit state relative to amount of phase fluctuations,
$\sigma$ and angle of conditioning detection, $\theta$. The values shown are as if verified by an imperfect detector. The parameters were
$V_x = 0.2$, $V_p = 2$ and $\eta = 0.85$.}
\label{varlimit3d}
\end{figure}
\begin{figure}[!t!]
\centerline{\psfig{figure=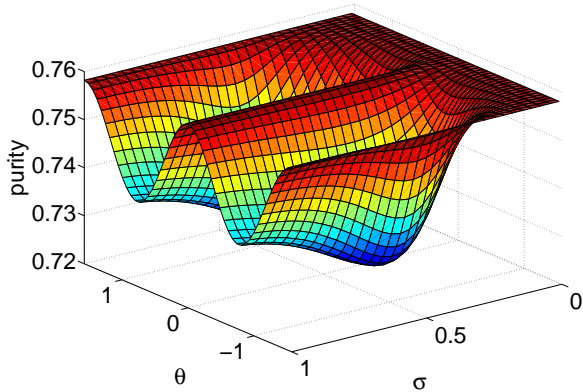,width=0.95\linewidth}}
\caption{\g{(Color online)} Purity of the asymptotic limit state relative to amount of phase fluctuations,
$\sigma$ and angle of conditioning detection, $\theta$. The values shown are as if verified by an imperfect detector. The parameters were
$V_x = 0.2$, $V_p = 2$ and $\eta = 0.85$.}
\label{purlimit3d}
\end{figure}

\jg{Some n}umerical results are shown in Fig.~\ref{limit}. It can be seen that the iterations
converge to the limit given by Eq. (\ref{Vlim}) in both \g{the} resulting variance and purity. Moreover, the purity
of the limit is independent \g{of} the actual amount of initial phase fluctuations and equals
to the purity of the original state (modified by imperfect detectors), although it is less
when the states are compared directly, without the imperfect verification measurement.


Figures \ref{varlimit3d} and \ref{purlimit3d} illustrate the behavior of the asymptotic limit
relative to \g{the} angle $\theta$ of \g{the} conditioning measurement. Although all choices of $\theta$
lead to a state purified to some extent, \g{the} two most interesting choices are $\theta = 0$,
corresponding to \g{the} measurement of \g{the} quadrature operator
$x$ and $\theta = \pi/2$, \g{corresponding to the} measurement of
$p$. Only these two choices can lead to \jg{a} purity of the final
state to be maximal, that is, equal to \g{the} purity of the initial state (modified by detection efficiency).
This is not very surprising, since these two choices coincide with the basis in which
 the covariance matrix of the initial state is diagonal.
We can see that
for weak phase noise (small $\sigma$) it is advantageous to employ conditioning on measurements
of \jg{the $p$ quadrature}  while for strong phase fluctuations  the  $x$ based post-selection is optimal.

\section{Summary}
We have provided \g{a} detailed analysis of multiple copy purification
\g{protocols} for phase diffused single-mode
squeezed states. We \g{performed} most calculations in the Fock basis, which allowed for
efficient numerical treatment. We have also discussed \g{the}
performance of collective purification, employing more than two copies
of the initial state in one step. The \emph{collective} purification/distillation uses an 
arbitrary number of copies to improve the purity and non-classicality of 
the state, and therefore was found to be a generalization of 
\emph{iterative} purification and distillation which requires $2^k$ copies. 
In particular we have shown that collective distillation based 
on three copies of phase diffused squeezed states is possible, and already 
provides a significantly improved degree of squeezing when compared with 
two-copy distillation.
We have also investigated the convergence of the iterative purification 
protocol and derived semi-analytical expressions for variances of the asymptotic Gaussian state. Finally, let us note that in the spirit of ref. \cite{nongdist1}, the results presented here can be straightforwardly extended to describe the distillation of phase diffused two-mode squeezed states.

\acknowledgments
P. M. acknowledges support from the European Social Fund.
J.F. acknowledges financial support from  the Ministry of Education of the Czech Republic
under the projects Centre of Modern Optics (LC06007) and Measurement and Information in Optics
(MSM6198959213), from GACR under project 202/07/J040 and  from the EU under project COVAQIAL
(FP6-511004). R.~S.~ acknowledges support from the Deutsche
  Forschungsgemeinschaft (DFG), project number SCHN 757/2-1.

\end{document}